\documentclass[final,5p, times, twocolumn]{elsarticle}
\usepackage{epsfig}
\usepackage{amssymb}
\usepackage{lipsum}
\usepackage{lineno}
\usepackage{graphicx}
\usepackage{xcolor}
\usepackage{hyperref}
\usepackage{comment}
\usepackage{scrextend}
\usepackage{amsmath}
\usepackage{mathtools}
\usepackage{newtxtext}
\usepackage{newtxmath}

\newcommand{\epem}{e$^+$e$^-$}
\newcommand{\epemmath}{\mathrm{e}^+\mathrm{e}^-}
\newcommand{\mee}{$M_{\mathrm{e}^+\mathrm{e}^-}$}
\newcommand{\meemath}{M_{\mathrm{e}^+\mathrm{e}^-}}

\newcommand{\mrhomath}{M_{\rho}}

\newcommand{\mmiss}{$M_{\rm{miss}}$}
\newcommand{\qdeux}{$q^2$}
\newcommand{\piz}{$\pi^0$}

\newcommand{\pim}{$\pi^-$}
\newcommand{\pimp}{$\pi^-$+ p}
\newcommand{\pippim}{$\pi^+\pi^-$}

\newcommand{\pimC}{$\pi^{-}+ \mathrm{C}$}

\newcommand{\chdeux}{CH$_2$}
\newcommand{\pimptonee}{\pimp ~$\to$~n~\epem }

\newcommand{\mevcc}{MeV/$c^{2}$}
\newcommand{\gevcc}{GeV/$c^{2}$}
\newcommand{\gevc}{GeV/$c$}

\newcommand{\gammas}{$\gamma^{\star}$}

\journal{Physics Letters B}

\makeatletter
\def\ps@pprintTitle{
   }
\makeatother

\begin{document}
\setcounter{page}{1}

%\linenumbers

\begin{frontmatter}

\title{First measurement of massive virtual photon emission from N* baryon resonances}

\author[6,14]{R.~Abou~Yassine}
\author[5]{J.~Adamczewski-Musch}
\author[10,9]{O.~Arnold}
\author[14]{E.T.~Atomssa}
\author[11]{M.~Becker}
\author[8]{C.~Behnke}
\author[10,9]{J.C.~Berger-Chen}
\author[1]{A.~Blanco}
\author[8,5,e]{C.~Blume}
\author[10]{M.~B\"{o}hmer}
% ...

\address[1]{LIP-Laborat\'{o}rio de Instrumenta\c{c}\~{a}o e F\'{\i}sica Experimental de Part\'{\i}culas, 3004-516~Coimbra, Portugal}
% ...
\address[5]{GSI Helmholtzzentrum f\"{u}r Schwerionenforschung GmbH, 64291~Darmstadt, Germany}
\address[6]{Institut f\"{u}r Kernphysik, Technische Universit\"{a}t Darmstadt, 64289~Darmstadt, Germany}
% ...
\address[8]{Institut f\"{u}r Kernphysik, Goethe-Universit\"{a}t, 60438 ~Frankfurt, Germany}
\address[9]{Excellence Cluster 'Origin and Structure of the Universe' , 85748~Garching, Germany}
\address[10]{Physik Department E62, Technische Universit\"{a}t M\"{u}nchen, 85748~Garching, Germany}
\address[11]{II.Physikalisches Institut, Justus Liebig Universit\"{a}t Giessen, 35392~Giessen, Germany}
% ...
\address[14]{Laboratoire de Physique des 2 infinis Irene Joliot-Curie, Universite Paris-Saclay, CNRS-IN2P3, F-91405~Orsay, France}
% ...
\address[c]{Also at Helmholtz Research Academy Hesse for FAIR (HFHF), Campus Darmstadt, 64390~Darmstadt, Germany}
% ...
\address[e]{Also at Helmholtz Research Academy Hesse for FAIR (HFHF), Campus Frankfurt, 60438~Frankfurt am Main, Germany}
% ...

\begin{abstract}
First information on the timelike electromagnetic structure of baryons in the second resonance region has been obtained from measurements of the dielectron (\epem ) invariant mass and angular distributions with the High Acceptance Di-Electron Spectrometer (HADES)  at GSI, using a secondary pion beam impinging on 
\chdeux\ and C targets. The quasi-free reaction \pimptonee\ at $\sqrt{s_{\pi \mathrm{p}}}$ = 1.49 GeV  is isolated in several analysis steps. The data set provides a crucial test for the description of baryon timelike transitions. Close to the kinematical limit the meassured \epem\ mass distribution shows deviations by up to a factor 8 from the Dalitz decay of a point-like baryon. Various theoretical approaches connecting information from hadronic and
electromagnetic channels in the spacelike and timelike regime  compare well with the measured \epem\ invariant mass distribution, including 
(i) a Vector Meson Dominance amplitude containing direct photon and vector meson ($\rho$) couplings to the baryon,
(ii) electromagnetic timelike baryon transition form factors in a covariant spectator-quark model, 
where meson cloud effects dominate, and (iii) application of dispersion theory, which demonstrates the importance of the pion electromagnetic form factors.

The dielectron angular distributions exhibit contributions of virtual photons ($\gamma^*$) with longitudinal polarization, in contrast to real photons. 
The $\gamma^*$ angular dependence supports the dominance of J=3/2, I=1/2 contributions  in both the $\gamma^{\star} n$ and the $\pi \pi $N channels.
\end{abstract}

\begin{keyword}
pion-nucleon interaction \sep dilepton \sep baryon Dalitz decay \sep off-shell $\rho$ meson \sep electromagnetic timelike transitions
%% PACS codes here, in the form: \PACS code \sep code
%% MSC codes here, in the form: \MSC code \sep code
%% or \MSC[2008] code \sep code (2000 is the default)
\end{keyword}
\end{frontmatter}

{\textbf {\textit {Introduction.}}}
Hadrons are composite quantum objects, characterized by a number of parameters, such as mass, charge, spin or form factors, which derive from the strong-interaction. However, despite the impressive progress made by first-principle approaches, concerted efforts of theory and experiment are still needed before  hadron properties can be  inferred directly from  Quantum Chromodynamics (QCD).  
This motivates, in particular, extensive studies of electromagnetic Transition Form Factors (eTFF) for the transition between a nucleon (N) and baryon resonances (R) (N\gammas\  $\rightarrow $ R) with electron scattering. In such reactions the squared four-momentum \qdeux\ of the virtual photon $(\gamma^*)$ is negative, probing the form factors in the spacelike region \cite{Aznauryan09,Thompson2001,Aznauryan13,Mokeev2016,Isupov17}.
However, a consistent understanding of baryon electromagnetic structure also requires experimental information in the timelike region \qdeux\ $> 0$ which is very scarce \cite{Berezhnev76_SovJourn24,Blokhintseva82,Hoffman83,Jerusalimov18}.
eTFF, which encode the electromagnetic structure of the involved baryons,  are analytical functions of \qdeux , which are closely related to helicity amplitudes. These functions are connected in the spacelike and timelike regions by dispersion relations.
The results obtained for various baryon transitions in the spacelike region reveal contributions from a quark core and a meson cloud, being an important feature at small \qdeux\   \cite{JuliaDiaz08,Aznauryan17}. The timelike  kinematic domain  can be addressed via the Dalitz decay of baryons R $\to$  N\epem , which probe electromagnetic transitions in the \qdeux = $M^2_{\mathrm{e}^+\mathrm{e}^-}$  interval [$4m_\mathrm{e}^2,(M_\mathrm{R}-M_\mathrm{N})^2$]. Here  \mee\ is the dielectron invariant mass and $m_\mathrm{e}$, $M_\mathrm{R}$ and $M_\mathrm{N}$ are the electron, resonance and nucleon masses, respectively.   
According to Vector Meson  Dominance (VMD) models, light vector mesons (VM = $\rho$ and $\omega$) should play an important role in these transitions. A strong enhancement of eTFF as a function of \mee\ approaching VM poles is expected. Indeed,  such effects have been observed in meson Dalitz decays, \cite{Berghauser2011,Aguar-Bartolome2013,Arnaldi2016,Adlarson2017} but confirmation is still lacking for baryons. 

The interest in  virtual photon coupling to baryons is also driven by the quest for $\rho$ meson  in-medium modifications, extensively investigated in heavy-ion collisions by means of dilepton spectroscopy (for a review, see \cite{Salabura2020}). In the context of VMD, the observed strong broadening of the $\rho$ spectral function is explained mainly by $\rho$-baryon resonance interactions \cite{Rapp2000}, with a major contribution of N(1520) in the second resonance region \cite{Peters98}. However, the mechanism of the coupling is not precisely known.
As first pointed out in \cite{Friman:1998fb}, the VMD model assuming that the coupling proceeds only via $\rho$ (VMD2) strongly overestimates R $\rightarrow$ N$\gamma$ branching ratios while this is avoided in the version of Kroll and Zummino \cite{Kroll67} (VMD1) with a superposition of direct photon and $\rho$ meson coupling. It allows to take into account the apparent non-universality of hadron couplings to conserved currents \cite{Oconnell95,Brown:1985gu} and to fix separately the $\rho$N and $\gamma$N branching ratios.  Consequently, in calculations of the in-medium $\rho$ spectral function of \cite{Rapp2000} a two-component scheme was applied. 

Dedicated microscopic calculations for the \pimptonee\ \cite{Lutz03,Zetenyi20} reaction were performed using effective Lagrangian models and the VMD described above. Covariant quark models \cite{Ramalho17,Ramalho20}, taking into account the coupling of the photon to the constituent quark core and to the meson cloud, have been recently extended to the timelike region. The results reveal the dominant role of the latter, with the $\rho$ meson  saturating the transition.
Effective field, dispersion theory, Dyson-Schwinger  or lattice QCD approaches  also offer promising perspectives for the calculations of eTFF of baryons \cite{Leupold2012,  Granados17cib, Eichmann16, Miramontes2021}. In particular, a very recent calculation based on the dispersion theory provides numerical predictions for the N-N(1520) transition \cite{AnDi2024}.
 
After the very first determination of the $\Delta (1232)$ Dalitz decay branching ratio   \cite{Hades17_DeltaDalitz} and investigations of higher lying resonances  \cite{Agakishiev12_pp35,Agakishiev12_pp22} using p + p reactions, the HADES collaboration started new measurements  in the second resonance region, using the GSI pion beam facility \cite{Hades17_pibeam,HadesPionbeamTwopi}. As  baryon resonances can be excited in pion induced reactions with a fixed mass in the $s$-channel, they are indeed ideally suited for  baryon Dalitz decay studies and offer a unique access to the baryon timelike electromagnetic structure, via the measurement of \epem\ invariant mass and angular distributions.  The \pimptonee\ reaction, often referred to as Inverse Pion Electroproduction, has been previously studied either at rest or at energies up to 400 MeV/c (see \cite{Surovtsev2005} and references therein), where only the $\Delta(1232)$ resonance contributes. Our data meet the widely expressed demand for new measurements of this reaction \cite{Briscoe2015,Briscoe2023} and constitute the first investigation of the time like electromagnetic structure of baryons in the second resonance region.
Analysis and simulation inputs are described in detail in \cite{jointPRC}, where  results for the inclusive \epem\ production are discussed.  In this letter,  measurements of the quasi-free \pimptonee\ reactions  at a  center-of-mass energy $\sqrt{s_{\pi \mathrm{p}}}$ = 1.49 \gevcc\ using a polyethylene (\chdeux ) and a carbon (C) target are presented.    Data are compared with predictions of four models where eTFF are taken into account using the two-component VMD1 scheme \cite{Zetenyi20,jointPRC}, a covariant quark model \cite{Ramalho17,Ramalho20} and a dispersion theory approach \cite{AnDi2024}. 

{\textbf {\textit {Experiment.}}} The experiment was performed using a secondary  pion beam with  central momentum of 0.685~\gevc\  (momentum spread of $\pm$1.7$\%$ and incident flux of  10$^5$ pions/s  \cite{Hades17_pibeam}).
\chdeux  and C targets were used in separate runs \cite{jointPRC}.
HADES \cite{Agakichiev09_techn}  consists of six identical sectors covering polar angles between 18$^{\circ}$ and $85^{\circ}$ with respect to the beam axis.
 The information provided by the hadron-blind RICH detector and the tracking and time-of-flight systems  was used to reconstruct the four-momenta of the e$^+$ and e$^-$  \cite{Agakichiev09_techn, Sellheim_PhD}.
Details about track selection cuts,  combinatorial background (CB) subtraction, efficiency corrections as well as normalization of the measured yields can be found in \cite{jointPRC}. 
For the exclusive channels presented in this analysis, the signal to CB ratio is larger than 50, except for \mee\ $\in [100,140]$ \mevcc , where values close to unity are reached due to real photon  conversion following \piz\ decays.

{\textbf {\textit {Simulations.}}} As explained in detail in \cite{jointPRC}, three main sources of \epem\ pairs are expected to contribute in the energy range of our experiment and were accordingly implemented in simulations: (i) the \piz\ Dalitz decay \piz $\to \gamma$\epem ,   (ii)  the $\eta$ Dalitz decay  $\eta \to \gamma$\epem  , and (iii)  the exclusive \pimptonee\ reaction.   Interactions with carbon nuclei were treated  using a quasi-free participant spectator model with a cross section ratio $\sigma_{\mathrm{C}}/\sigma_\mathrm{p}$~=~2.9, consistent with the measured yields on the polyethylene and carbon targets \cite{jointPRC}. The comparison to the results of the inclusive channels allowed to validate the  description of the $\pi^0$ and $\eta$ sources.

For the exclusive \pimptonee\ reaction channel, isolated as described below, a ``QED reference" corresponding to the Dalitz decay of point-like baryons was first estimated in \cite{Landsberg1985}. The calculation was made for the N(1520) 3/2$^-$ and N(1535) 1/2$^-$ resonances, which are expected to be the dominant contribution and exhibit the same mass dependence of the partial Dalitz decay width (for details, see sec.~IV.A of \cite{jointPRC}). The normalization of the calculation is provided by the cross section of the \pimp $\leftrightarrow \gamma$~n reaction at the same center-of-mass energy.  This QED reference, with a total cross section of 2.14 $\mu$b, constitutes the basis for the quantification of the mass dependence of eTFF (see section IV.A of \cite{jointPRC} for details).  

To estimate the effect of the unknown \mee\ dependent eTFF, two models of eTFF were used in the simulations (for details see \cite{jointPRC}).
The first one is the two-component VMD1 model  introduced above \cite{Rapp2000,Kroll67,Oconnell95,Zetenyi20}.  
The $\rho$ contribution was  extracted in the Partial Wave Analysis (PWA) of the \pimp $\rightarrow$ n \pippim\ reaction measured in the same experiment  \cite{HadesPionbeamTwopi}.
The respective $ \Gamma_{\epemmath}(M_{\epemmath})$ vanishes at $\meemath = 0$ and follows a linear $\meemath$ dependence:
 $ \Gamma_{\epemmath}(M_{\epemmath}) = \Gamma_{\epemmath}(\mrhomath)\ \meemath\ / \mrhomath $, 
where $\Gamma_{\epemmath}(\mrhomath)$ is the partial decay width at the  Breit-Wigner mass $\mrhomath$. The spectral function has the form:
\begin{eqnarray}
 \mathcal{A}(M) \sim   \frac{M^2\Gamma_{\mathrm{tot}}(M)}{(\mrhomath ^2 - M^2)^2 + M^2\Gamma^2_{\mathrm{tot}}(M)}.
\label{eq:BW2}
\end{eqnarray}
The total $\rho$ decay width $\Gamma_{\mathrm{tot}}(M)$ includes a parametrization of the $\rho \to \pi \pi$ decay consistent with the PWA calculations (see \cite{jointPRC} for details).
The total dielectron yield is obtained as a coherent sum of the photon (``QED reference") and the $\rho$ meson contributions, derived as described above. The phase between the two contributions was considered as a free parameter.

In an alternative approach, an eTFF model is implemented, based on the recently developed covariant quark model  for the N-N(1520)~\cite{Ramalho17} and  N-N(1535) \cite{Ramalho20} transitions. 
The differential Dalitz decay width of 3/2$^{-}$ or 1/2$^{-}$ baryons depends on the eTFF via the expression
\begin{align}
\frac{1}{\Gamma^{\mathrm{N^{*}}\rightarrow \mathrm{N}\gamma}}\frac{d\Gamma^{\mathrm{N}^{\star} \rightarrow \mathrm{N} \mathrm{e}^+\mathrm{e}^-}}{d\meemath} =\frac{2\alpha}{3\pi M_{\mathrm{e}^+\mathrm{e}^-}} \frac{\sigma_{+}^{3}\sigma_{-}}{m_{+}^{3}m_{-}}  \frac{\left|G_\mathrm{T}(q^2)\right|^2}{\left|G_\mathrm{T}(0)\right|^2},  	
\label{eq:effectiveFF}
\end{align}
where            $m_\pm   =  M_{\mathrm{R}} \pm M_\mathrm{N} $ and 
$  \quad \sigma_{\pm}^2  =  m_\pm^2-q^2$. $\alpha$ is the fine-structure constant,  $\Gamma^{{\mathrm{N}^{*}\rightarrow \mathrm{N}\gamma}}$ is the radiative decay width and 
 $\left|G_\mathrm{T}(q^2)\right|^2$ is a combination of the squared moduli of the electric, magnetic  and Coulomb form factors, as given in \cite{jointPRC}.  
In our simulation,  the mass dependence of  $\left|G_\mathrm{T}(q^2)\right|^2$ is taken from the model of \cite{Ramalho17,Ramalho20} , in which it is governed by the pion form factor and is similar for N(1520) and N(1535). The overall normalization of the calculation was adjusted to ensure consistency of 
 the dielectron yield with the \pimp $\to$ n$\gamma$ cross section.
 
\begin{figure}[!hbt]
	\centering
	\includegraphics[width=0.4\textwidth]{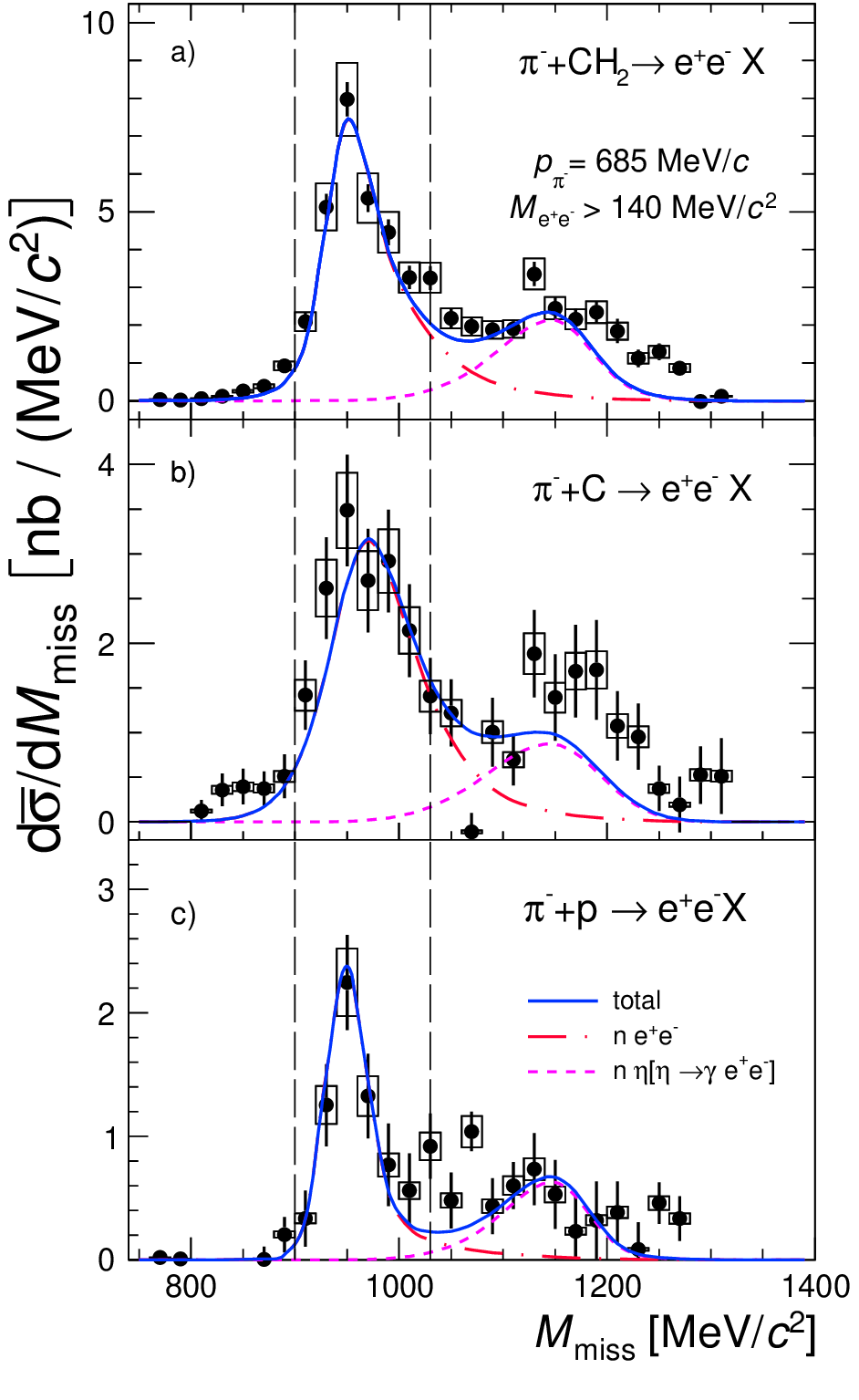}
		\caption{Distribution of the missing mass \mmiss\ calculated $w.r.t.$ the \pimp\ system for a fixed pion beam momentum of 0.685 \gevc\ for events  with an \epem\ invariant mass larger than 140 \mevcc. Data (symbols with statistical errors (bars) and  systematic errors (boxes))   for \pim + \chdeux\ (a), \pim + C (b) and \pim + p (c) are displayed  after efficiency corrections. Simulated distributions (curves) after full reconstruction  are shown for the  \pimptonee\ channel using the eTFF model (dotted-dashed red), $\eta$ Dalitz decay (dashed magenta) and the total (blue). The vertical dashed lines indicate the missing mass window used to select events of the \pimptonee\ reaction.}
	\label{fig:MM}
\end{figure}

{\textbf {\textit {Exclusive quasi-free channel selection.}}} For \mee\ larger than the \piz\ mass, the selection of the quasi-free \pimptonee\ channel is based on the missing mass $M_{\mathrm{miss}}=\sqrt{(P_{\mathrm{in}}-P_{\mathrm{e}^+\mathrm{e}^-})^2}$, determined by the four momenta  $P_{\mathrm{in}}$ and $P_{\mathrm{e}^+\mathrm{e}^-}$   of the initial \pim p and final \epem\ systems, respectively.
$P_{\mathrm{in}}$ is calculated using a fixed pion momentum of 0.685 \gevc\ and assuming a free target proton at rest.
The cross sections integrated over the HADES acceptance (denoted by $\bar{\sigma}$ throughout this letter) for the  C target (Fig.~\ref{fig:MM}b)  have been subtracted from that of  the \chdeux\ (Fig.~\ref{fig:MM}a) to determine the distribution for \pimp\ interactions (Fig.~\ref{fig:MM}c).
The three distributions show two similar structures, which are well reproduced by the simulations: a peak around the neutron mass, which can be attributed to the exclusive \pimptonee\ reaction, either on a free proton or a bound proton in the carbon nuclei and a broader structure, primarily due to the $\eta$ Dalitz decay, as validated in \cite{GallmeisterPhysRevC.106.064910}.
The width of the neutron peak obtained for the free \pimptonee\ reaction is mainly due to the detector resolution, with only a minor impact of the pion beam momentum spread,  while, for the carbon target, it mainly results from the Fermi motion of the nucleon in the carbon nucleus. 
It can also be observed that the eTFF model, which is used here to describe the \pimptonee\ process, provides a good description of the yields. 
The second structure is well reproduced by the simulation of the $\eta$ Dalitz decay, although contributions from the Bremsstrahlung process, of the type \pim N $\to  \pi $N \epem, might be responsible for the missing yield at $ M_{\rm{miss}} > M_N + m_\pi$.  These results are fully consistent with the analysis of the inclusive \epem\ production \cite{jointPRC}.
Due to the low statistics recorded on the carbon target (Fig.~\ref{fig:MM}b), the precision obtained separately for \pimp\ and \pimC\ interactions is limited and does not allow for an accurate study of the corresponding distributions as a function of invariant mass or angles. One should stress, however, that the very good description of the neutron peak in the \pimC\ case justifies the assumption about the quasi-free character of the reaction.  
Therefore, the focus of this analysis is on the polyethylene target measurements, applying the missing mass selection $900$ \mevcc $ <$ \mmiss $ <  1030 $~\mevcc\ to enhance  the exclusive quasi-free \pimptonee\ channel. 

\begin{figure}[!hbt]
 \centering
  \includegraphics[width=0.48\textwidth]{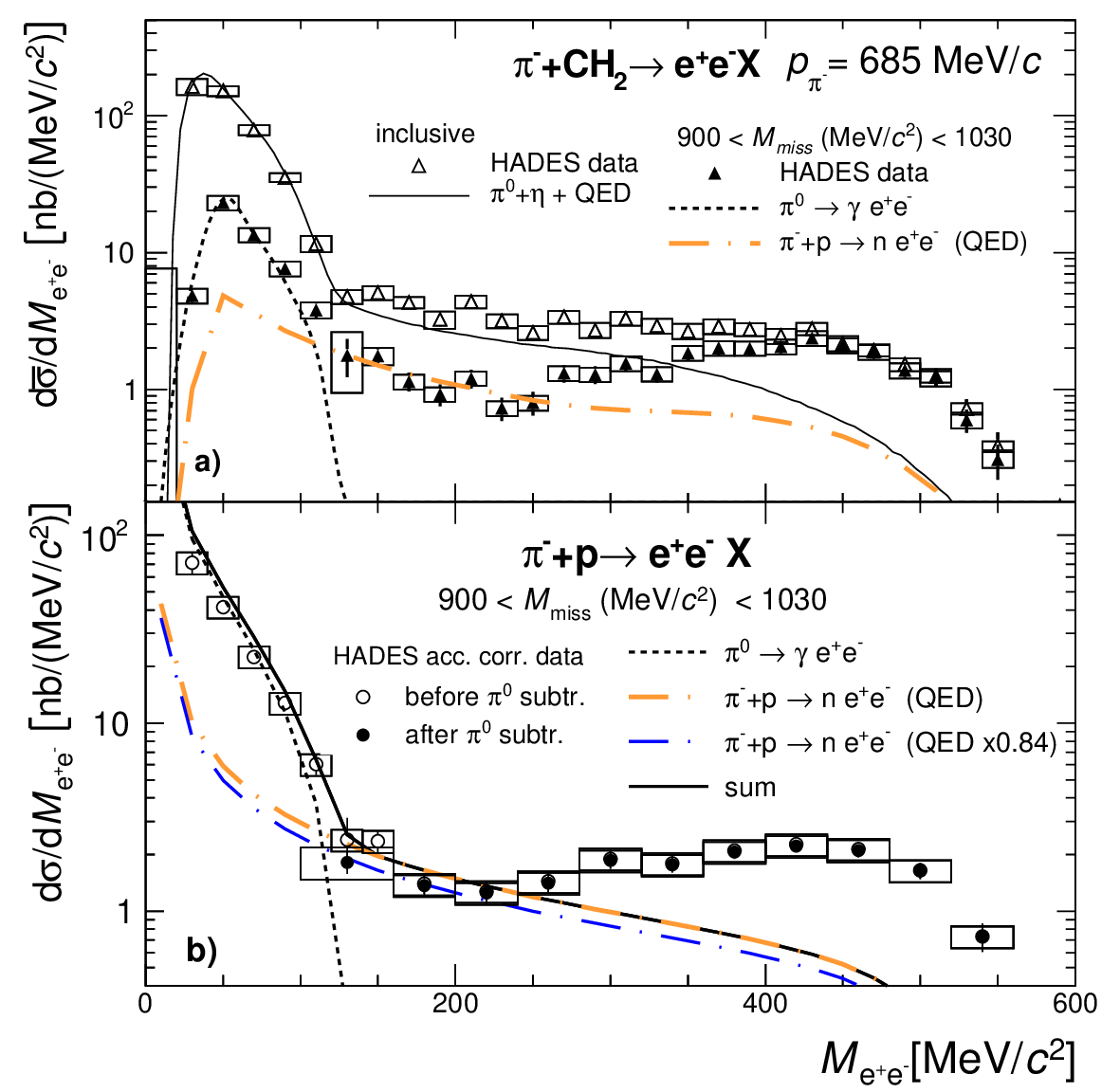}
\caption{(a)  d$\bar{\sigma}$/ d$\meemath$  for the \pim +\chdeux\ reaction integrated in the HADES acceptance. Open triangles: total yields, full triangles:  yields integrated over the missing mass range [900-1030~\mevcc ] . The full thin black curve displays the simulation including the \piz , $\eta$ and point-like baryon Dalitz  (``QED reference") decays. The black dotted and the orange dashed-dotted curve  depict the  \piz\ Dalitz decay and the QED reference, respectively, after applying the missing mass selection. (b) d$\sigma$/ d$\meemath$  for the  quasi-free \pimptonee\ reaction obtained from the \pim + \chdeux\ data integrated over the missing mass range [900-1030~\mevcc ]  obtained after normalization by the number of effective protons and acceptance corrections. Data  before and after  subtraction of the residual \piz\ contribution are shown as empty and full disks, respectively. The curves exhibit the simulation results for the different processes~: QED reference  (d$\sigma$/d $\meemath)_{\mathrm{QED}}$ (orange dashed-dotted), \piz\ Dalitz decay (dotted curve) and the sum (thin black curve, covering the QED curve above 150 \mevcc ). The blue dashed-dotted curve displays the renormalized QED curve used for the extrapolation of the cross sections  and for the calculation of the experimental eTFF.
        Symbols  show the data with statistical  and systematic point-to-point errors, as bars and boxes, respectively. Curves  display simulations  with absolute normalization in the two panels.}
	\label{fig:Minv_twopads}
\end{figure}

{\textbf {\textit {Invariant masses.}}} Figure~\ref{fig:Minv_twopads}a) shows the effect of the missing mass cut ($900$ \mevcc $ <$ \mmiss $ <  1030 $~\mevcc) on the cross sections integrated over the HADES acceptance (d$\bar{\sigma}$/ d$\meemath$) and the comparison to a simulation including the \piz , $\eta$ and point-like baryon Dalitz decays. The $\eta$ contribution is fully suppressed by the missing mass cut, but the \epem\ yield  below 100 \mevcc\ is still strongly dominated by the  \piz\ Dalitz decay. The simulation is used to subtract this contribution and extract the exclusive yield in the region between 100 and 140 \mevcc . Finally, 1280 \epem\ pairs with \mee $>$ 100 \mevcc\ corresponding to the quasi-free \pimptonee\ reaction are isolated.  The further analysis steps are illustrated in Fig.~\ref{fig:Minv_twopads}b. Acceptance correction factors are deduced from the simulation with the eTFF model, including also the \piz contribution for the data obtained before the \piz\ subtraction. This leads to an  average correction factor of about 5.9 with a smooth invariant mass dependence. Due to the model dependence of this correction, the resulting yields are affected by a systematic error of 5~$\%$.  
To determine  cross sections for the reaction \pimptonee, the  number of effective protons Z$_{\rm{eff}}+2 = 4.9\pm 0.5$ \cite{jointPRC}, contributing to the \pim + \chdeux\ reaction, has been taken into account. The statistical and point-to-point systematic errors are indicated in Fig.~\ref{fig:Minv_twopads}. In addition, a global systematic error of about 17\% has to be considered, due mostly to the uncertainties on the efficiency corrections  ($\pm$ 12\%) and on the effective proton number ($\pm$10\%).

\begin{figure}[!hbt]
 \centering
   \includegraphics[width=0.48\textwidth]{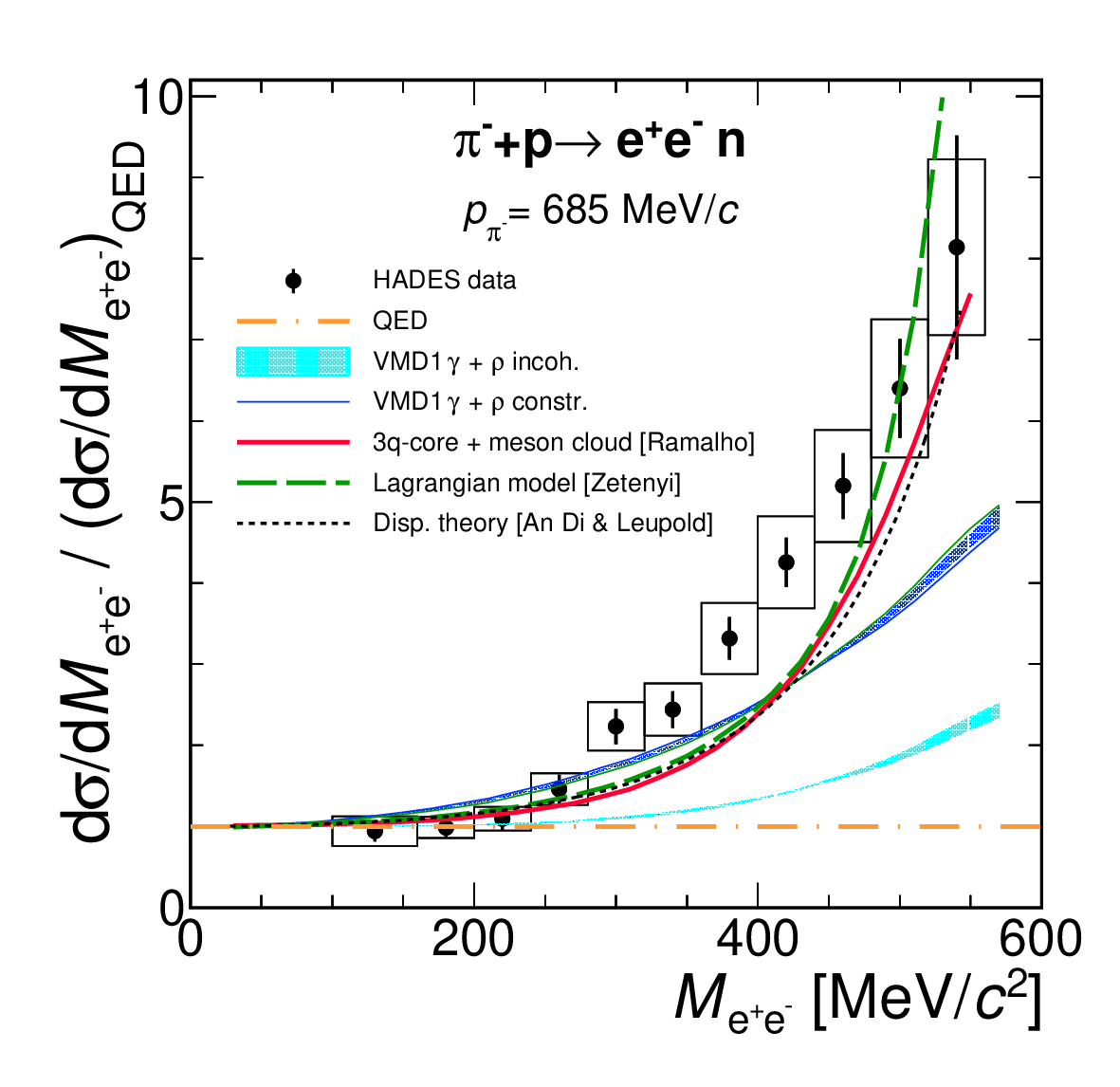} 
		\caption{Ratios (d$\sigma$/d$\meemath$)/(d$\sigma$/d$\meemath)_{\mathrm{QED}}$ deduced from the measurements (black dots)  are compared to model predictions. Statistical and systematic errors are shown as bars and boxes, respectively. In addition, the global uncertainty due to the QED normalization (see text) amounts to 17\%. Blue  (cyan) colored areas: two-component VMD model with constructive (incoherent) sum of $\rho$ and $\gamma$ contributions.  Calculations using the  timelike form-factor, the Lagrangian  and  the dispersion theory models are shown as red solid, green long dashed and dotted black curves, respectively. See text for details about the normalization of the QED reference for the data and Lagrangian models.}
	\label{fig:eTFF}
\end{figure}

In the  \mee range between 100 and 250 \mevcc , the experimental cross sections obtained after \piz\ subtraction  are in fair agreement with the QED reference derived from the \pimp $\leftrightarrow$ n$\gamma$ measurements and used in the simulations, as described above. It can still be noted that the measured cross sections are systematically lower by a factor 0.84 $\pm$ 0.14, where the uncertainty is mostly due to the point to point systematic errors. This renormalization of the QED reference, shown as a blue dashed-dotted curve in Fig.~\ref{fig:Minv_twopads}b, is consistent with the above-mentioned global uncertainty of HADES data of about 17\%.  However, at larger invariant masses, an excess of up to a factor 8 is observed, pointing to a strong effect of the eTFF.
To demonstrate this effect more directly, the ratios of data yields to the renormalized QED reference  are displayed in Fig.~\ref{fig:eTFF}, which is a measure of the effective eTFF of $N^{\star}$ baryons excited in the \pimptonee\ reaction.  

{\textbf {\textit {Comparison to models.}}} 
The various eTFF models mentioned above can be confronted with data. First, it is interesting to check the  VMD1 model.      As shown in Fig.~\ref{fig:eTFF}, the yields above 300 \mevcc\  depend  on the interference between the $\rho$ and $\gamma$ amplitudes. An incoherent sum of the two contributions (cyan colored area) significantly underestimates the data, but a better description of the \epem\ invariant mass spectrum can be obtained using a maximum constructive interference (blue colored area). As shown in detail in \cite{jointPRC}, the effect of the Fermi motion on the simulated invariant mass distribution for the quasi-free reaction is very small.

Simulations based on the eTFF model \cite{Ramalho17,Ramalho20} for these resonances (red curve in Fig.~\ref{fig:eTFF}) also give a satisfactory description of the data, which demonstrates  that the meson cloud contribution, driven by the pion electromagnetic form factor, plays indeed a dominant role.
Our data are also compared to the microscopic calculation of \cite{Zetenyi20} based on an effective Lagrangian approach, taking into account various resonant and non-resonant amplitudes in a coherent way. The calculation uses the  N$^{\star}$N$\rho$ couplings derived from the  PWA \cite{HadesPionbeamTwopi}.  A salient feature of this model is the application of the two-component VMD model to all baryon-photon couplings.   Choosing a relative phase of 90$^{\circ}$ between the resonant $\gamma$ and $\rho$ amplitudes, a good description of the \epem\ production is achieved, as shown by the green long dashed curve in Fig.~\ref{fig:eTFF}). For the calculation of the eTFF for this model, a QED reference renormalized by a factor 1.18 has been chosen, to compensate for the higher model predictions in the low invariant mass region. 
Finally,  our data are compared to  a new model for the N-N(1520) isovector eTFF based on the dispersion theory \cite{AnDi2024} and relying at low energy on the pion electromagnetic form factor and pion-baryon scattering amplitudes. Subtraction constants encoding short-distance physics and coupling constants for the three-point N(1520)-$\pi$-N and N(1520)-$\pi$-$\Delta$ interactions are fitted to spacelike form factors and hadronic decays. In particular, strong constraints are provided by the recent HADES measurements of the N(1520) decay to $\Delta$$\pi$ and N$\rho$ channels \cite{HadesPionbeamTwopi}. However, due to a residual sign ambiguity for the N(1520)$-\pi$-N  coupling, two sets of parameters are obtained.   Model predictions do not depend much on these parameters for M$<$ 300 \mevcc , but deviations by up to a factor 2 can be observed at the highest invariant masses (see \cite{AnDi2024} for details). The eTFF values, calculated  for the central value of the parameter set 2 \cite{AnDi2024} , using our QED reference, are shown as a dotted black curve in Fig.~\ref{fig:eTFF}. A fair description of the data is obtained. Although our analysis does not allow to isolate the N(1520) contribution, this result is a first stimulating step towards a global dispersive approach of various electromagnetic and hadronic  channels.

The measured  \epem\ cross section for \mee ~$\approx$~500~\mevcc\  is more than two orders of magnitude larger than the calculations of  \cite{Lutz03}, which were based on a very low off-shell $\rho$  cross section and strong destructive interferences with off-shell $\omega$ production. The calculations of \cite{Titov01}, which were performed for $\sqrt{s_{\pi \mathrm{p}}}$ larger than 1.6 GeV, also predicted large negative interferences between $\rho$ and $\omega$, though with a larger $\rho$ yield.  

The renormalized ``QED reference" model was used to extrapolate the experimental differential cross section  to small invariant masses ($M_{\mathrm{e}^{+}\mathrm{e}^{-}}< 120$  MeV/c$^2$). In addition, the reduction of the yield due the missing mass cut, which is of the order of 18\% and slightly depending on the invariant mass, was corrected for.  In this way, a total  cross section of  $\sigma  = (2.25 \pm 0.06 ^{\, \mathrm{stat}} \pm 0.47^ {\, \mathrm{syst}})\  \mu b $ is estimated for the \pimptonee\ reaction, where the important sources of uncertainties originate from the efficiency correction ($\pm$12\%), the effective number of protons ($\pm$10\%) and the normalization of the QED reference used for the extrapolation at small \mee ($\pm$10\%).
The ratio of the integrated experimental and renormalized “QED reference” cross sections, which, as explained above, quantifies the effect of the effective eTFF on the \pimptonee\  cross section amounts to $1.26  \pm 0.03 ^{\,\mathrm{stat}} \pm 0.10^{\, \mathrm{syst}}$. Here, the dominant source of systematic error is the extrapolation, as the other contributions mentioned above mostly cancel in the ratio. 

{\textbf {\textit {Angular distributions.}}}
\begin{figure}[hbt]
  \begin{center}
 \includegraphics[width=0.4\textwidth]{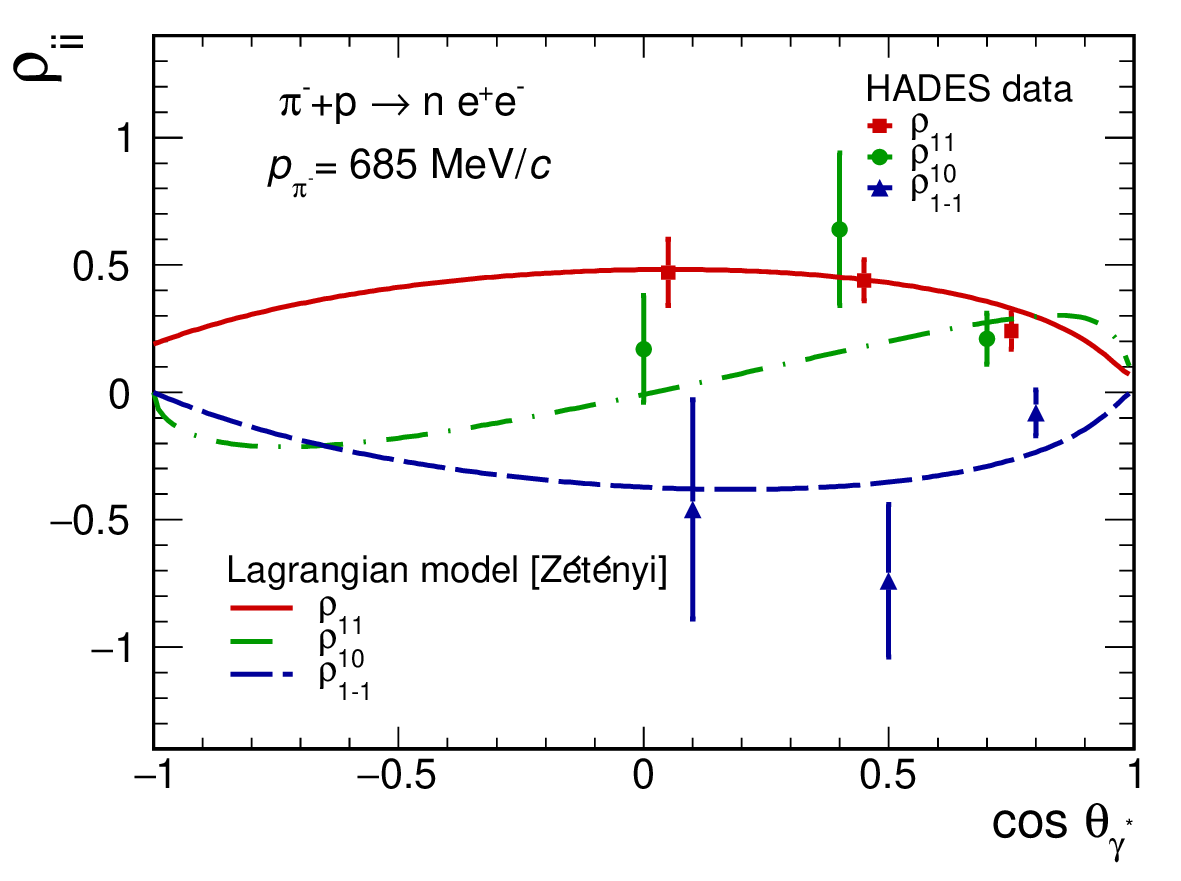} 
   \caption{Results of the analysis of   the $e^+$ and $e^-$  angular distributions  in the quasi-free \pimptonee\ reaction at $\sqrt s_{\pi p}=1.49 $ GeV. Spin density matrix elements $\rho_{ij}$: $\rho_{11}$ (red symbols), $\rho_{10}$ (green symbols) and  $\rho_{1-1}$ (blue symbols) extracted for \epem\ events  with \mee$ > 300$ \mevcc , are shown in 3 bins of the cosine of the virtual photon angle $\theta_{\gamma^*}$ ([-0.2,0.3], [0.3,0.6], [0.6,0.95]) compared to the predictions of the  model \cite{Zetenyi20} for $\rho_{11}$ (full curve), $\rho_{10}$ (dash-dotted curve) and  $\rho_{1-1}$ (dashed curve). The experimental uncertainties, which are dominated by statistical errors, are displayed as bars. 
}
  \label{fig:SpinDensityMatrix}
  \end{center}
\end{figure}
 Further information on the nature of the baryon timelike electromagnetic transitions in the \pimptonee\ reaction  can be obtained from the angular distributions.   A convenient parameterization of the differential cross sections  $d^4 \sigma /  d \Theta_{\gamma*} d M_{e^+ e^-} d \cos \Theta \, d \phi \propto \vert A \vert^2 $
is provided by the density matrix formalism  \cite{Zetenyi20,Sarantsev18,Speranza17} with the relevant dependencies of the squared modulus  $\vert A \vert^2$ of the amplitude at given value of \mee\ and polar angle ($\Theta_{\gamma*}$) of the virtual photon in the center-of-mass frame:
\begin{eqnarray}
&& \vert A \vert^2 
 \propto   8 k^2 \bigl[1 - \rho_{11} 
 + \left( 3\rho_{11} -1 \right)\cos^2\Theta \nonumber \\ && 
+ \sqrt{2} \mathrm{Re} \rho_{10} \sin2\Theta \cos\phi	+ \mathrm{Re} \rho_{1-1}\sin^2\Theta \cos2\phi \bigl].
\end{eqnarray}
Here,  $k$, $\Theta$ and $\phi$ denote the momentum, the polar and azimuthal angles of the electron in the virtual photon reference frame, respectively, and $\rho_{11,10,1-1}$ are the three independent  density matrix coefficients, which for a given transition are related to the corresponding eTFF \cite{Zetenyi20,Speranza17}. 
A method to extract these quantities from a fit to experimental data has been developed taking into account acceptance and detector inefficiency effects \cite{Sarantsev18,Ramstein:2020jkr,Scozzi18}. 
Although the statistics of the measurement is limited,  the coefficients can be extracted as a function of the \gammas\ emission angle in the center-of-mass   for \mee $>$ 300~\mevcc , as shown in Fig.~\ref{fig:SpinDensityMatrix}.    
The significant deviations of  $\rho_{11}$  from 0.5 and  of $\rho_{10}$  from 0  observed for the last  angular bin clearly demonstrate the contributions of virtual photons with longitudinal polarization, in contrast to real photons. 
The angular dependence of the coefficients indicate  important contributions of spins  larger than 1/2, in agreement with the dominance of the N(1520) 3/2$^-$ resonance in the $\rho$ production, which was found in the PWA of the two-pion production \cite{HadesPionbeamTwopi}.
The results are also compared to the predictions of the Lagrangian model \cite{Zetenyi20} for an invariant mass \mee = 400~\mevcc\ and  a phase $\Phi=90^{\circ}$ between the photon and $\rho$ contribution, to be consistent with the comparison of eTFF (Fig.~\ref{fig:eTFF}).  The  main trend of the data is accounted for, which suggests that the model describes realistically  the virtual photon polarization and the dominant  J=3/2 contribution.

{\textbf{\textit{Summary.}}} In summary, \epem\ production has been measured   in pion induced reactions at a momentum of 0.685 \gevc , opening a window on the timelike electromagnetic structure of baryon transitions in the second resonance region. The data  for the \pimptonee\ reaction for invariant masses below 250~\mevcc\ is consistent with the \pimp ~$\rightarrow \gamma$~n results  while an excess by up to a factor 8 is observed for larger invariant masses. 

This effect is qualitatively described, albeit with smaller eTFF values by up to 40\%, within a two-component model, based on the coherent sum of a direct photon
and a $\rho$ amplitude derived from the VMD1 Lagrangian. Making use of the existing PWA solution for the two-pion production at the same energy \cite{HadesPionbeamTwopi}, this approach provides a simple data-driven description of electromagnetic and hadronic channels in the \pimp\ reaction. A Lagrangian model based on the same VMD1 approach for resonant and non-resonant contributions \cite{Zetenyi20},
is  in fair agreement with the experimental invariant mass distribution. Both models support the two-component  character of the timelike electromagnetic baryon transition, with the coupling via vector mesons taking over the direct coupling with increasing invariant mass. \par
Our data are also consistent with the predictions of a covariant  eTFF model of N-N(1520) and N-N(1535)   \cite{Ramalho17,Ramalho20}. Within this model, the enhancement is explained by a strong meson cloud contribution, in line with conclusions derived from electron scattering experiments in the spacelike region.  \par
Last but not least, our results provide a very novel benchmark test for recent models deriving from dispersion relation theory \cite{AnDi2024}, opening the exciting perspective of a description of hadronic and electromagnetic channels in a common scheme, grounded in  first principles. \par
The successful description of the data in the whole invariant mass range by  models, which derive from very different approaches is noticeable. This can be attributed to the fact that they  all take into account, on the one hand, the information available in the spacelike, or at least the photon-point region and on the other hand  the $\rho$ contribution, either explicitly or via the pion electromagnetic form factor. For the first time, our experimental results quantify the important role of the off-shell $\rho$ meson in baryon resonance electromagnetic transitions. \par

The present results do not yet allow to separate non-resonant and resonant contributions  in a model independent way.  The nature of the transition, with a dominant role of the   N-N(1520),  is however corroborated by the extraction of the spin density matrix elements, which are well described by the Lagrangian model.  On the other hand,  the information on the virtual photon polarization is limited due to the large experimental uncertainties.  \par
 
The pioneering results presented in this letter  constitute a new insight into the nature of baryon-virtual photon coupling in the timelike region and allow for a test of very different theoretical approaches, including modern models based on dispersion theory. 
Future experiments aiming for high statistics scans of the third and second resonance regions combined with measurements of  the dielectron invariant masses and angular distributions  are planned by the HADES collaboration. The on-going theoretical efforts will be beneficial  to extract further information on the baryon timelike electromagnetic structure.

\section*{Acknowledgments}
We would like to warmly thank An Di and Stefan Leupold for providing results from their very recent calculations. We appreciated the longstanding support and the theoretical tools  provided by A. Sarantsev, G. Ramalho and M.~T. Pe\~na.
 We are very grateful for interesting discussions on theoretical aspects with K. Gallmeister and U. Mosel.
We acknowledge support from SIP JUC Cracow, Cracow (Poland), National Science Center, 2016/23/P/ST2/04066 POLONEZ, 2017/25/N/ST2/00580, 2017/26/M/ST2/00600, SONATA-BIS 2023/50/E/ST2/00673, 2023/49/B/ST2/00652 (OPUS); WUT Warszawa (Poland) No: 2020/38/E/ST2/00019 (NCN), IDUB-POB-FWEiTE-3; TU Darmstadt, Darmstadt (Germany), DFG GRK 2128, DFG CRC-TR 211, BMBF:05P18RDFC1, HFHF (Campus Darmstadt), ELEMENTS 500/10.006, VH-NG-823,GSI F\&E, EMMI at GSI Darmstadt; Goethe-University, Frankfurt (Germany), BMBF:05P12RFGHJ, GSI F\&E, HFHF (Campus Frankfurt), EMMI at GSI Darmstadt, ELEMENTS 500/10.006; TU Munchen, Garching (Germany), MLL Munchen, DFG EClust 153, GSI TMLRG1316F, BMBF 05P15WOFCA, SFB 1258, DFG FAB898/2-2; JLU Giessen, Giessen (Germany), BMBF:05P12RGGHM; IJCLab Orsay, Orsay (France), CNRS/IN2P3, P2IO Labex, France; NPI CAS, Rez (Czech Republic), MSMT LM2023060, MSMT OP JAK CZ.02.01.01/00/23\_015/0008181. \\
\\
We acknowledge the contribution from the following colleagues: A. Belyaev, O. Fateev, M. Golubeva, F. Guber, A. Ierusalimov, A. Ivashkin, A. Kurepin, A. Kurilkin, P. Kurilkin, V. Ladygin, A. Lebedev, S. Morozov, O. Petukhov, A. Reshetin, A. Sadovsky

%TC:endignore
%

\bibliographystyle{elsarticle-num} 
\bibliography{dilepton_pionBeam}
\end{document}